\documentclass[twocolumn,prl,aps,superscriptaddress, longbibliography]{revtex4-1}
\usepackage[T1]{fontenc}
\usepackage[latin9]{inputenc}

\usepackage{amsmath}
\usepackage{amssymb}
\usepackage{xcolor}
\usepackage{bbm}
\usepackage{braket}
\usepackage{mathrsfs}  
\allowdisplaybreaks

\usepackage{comment}

\usepackage{graphicx}

\usepackage[colorlinks=true]{hyperref}  
\hypersetup{
    bookmarks=true,         
    unicode=false,          
    pdftoolbar=true,        
    pdfmenubar=true,        
    pdffitwindow=false,     
    pdfstartview={FitH},    
    pdftitle={Composite fermions},    
    pdfauthor={Ruegg, Chaudhary, Slager},     
    pdfsubject={},   
    pdfcreator={},   
    pdfproducer={}, 
    pdfkeywords={} {} {}, 
    pdfnewwindow=true,      
    colorlinks=true,       
    linkcolor=blue, 
    citecolor=blue,        
    filecolor=magenta,      
    urlcolor=blue,           
	breaklinks=true
}

\definecolor{orange}{rgb}{1,0.5,0}

\newcommand{\sect}[1]{\vspace{0.3em}{\it #1.}---}

\usepackage{bbold}

\newcommand{\be}{\begin{equation}}
\newcommand{\ee}{\end{equation}}
\newcommand{\bea}{\begin{eqnarray}}
\newcommand{\eea}{\end{eqnarray}}

\renewcommand{\vec}[1]{\boldsymbol{#1}}

\newcommand{\ie}{{\it i.e.}~}

\newcommand{\cf}{{\it c.f.}~}

\newcommand{\cd}{c^{\dagger}}
\newcommand{\hd}{h^{\dagger}}

\newcommand{\psid}{\psi^{\dagger}}

\newcommand{\chid}{\chi^{\dagger}}

\begin{document}

\title{Pairing of Composite-Electrons and Composite-Holes in $\nu_T=1$ Quantum Hall Bilayers}

\author{Luca R\"uegg}
\email{lr537@cam.ac.uk}
\address{TCM Group, Cavendish Laboratory, University of Cambridge, J. J. Thomson Avenue, Cambridge CB3 0HE, United Kingdom}
\author{Gaurav Chaudhary}
\email{gc674@cam.ac.uk}
\address{TCM Group, Cavendish Laboratory, University of Cambridge, J. J. Thomson Avenue, Cambridge CB3 0HE, United Kingdom}
\author{Robert-Jan Slager}
\address{TCM Group, Cavendish Laboratory, University of Cambridge, J. J. Thomson Avenue, Cambridge CB3 0HE, United Kingdom}

\date{\today}
\begin{abstract} 
Motivated by recent experimental indications of preformed electron-hole pairs in $\nu_T=1$ quantum Hall bilayers at relatively large separation, we formulate a Chern-Simons (CS) theory of the coupled composite electron liquid (CEL) and composite hole liquid (CHL). 
We show that the effective action of the CS gauge field fluctuations around the saddle-point leads to stable pairing between CEL and CHL.  We find that the CEL-CHL pairing theory leads to a dominant $s$-wave channel in contrast to the dominant $p$-wave channel found in the CEL-CEL pairing theory.
Moreover, the CEL-CHL pairing is generally stronger than the CEL-CEL pairing across the whole frequency spectrum.
Finally, we discuss possible differences between the two pairing mechanisms that may be probed in experiments.
\end{abstract}
\maketitle

\sect{Introduction}
Quantum Hall (QH) bilayers exhibit many phenomena that are absent in the QH monolayers~\cite{Eisenstein1992,Eisenstein2014}. 
At the total filling fraction $\nu_T = \nu_+ + \nu_- = 1$ ($+ / -$ refer to the upper/lower layer), when the two layers have almost equal carrier densities and small layer separation $d$ (characterised by the ratio: $d/\ell$, where $\ell = \sqrt{\hbar/(eB)}$ is the magnetic length), the system exhibits a remarkable exciton condensate (XC) phase~\cite{Spielman2000,Eisenstein2004}. 
This superfluid phase of excitons was first observed in a Josephson-like zero-bias peak of the tunneling current between the two layers~\cite{Spielman2000}, and was later confirmed by a vanishing counterflow Hall resistance~\cite{Kellogg2004} and a perfect Coulomb drag~\cite{Nandi2012}. 

In the $d \rightarrow 0$ limit, the system can be viewed as a two-component monolayer QH system, which is an incompressible state described by the Halperin $(1,1,1)$ wavefunction~\cite{Halperin1983}. 
The enriched physics associated with the layer degree of freedom is manifested in a broken symmetry~\cite{Fertig1989} and spontaneous interlayer phase coherence~\cite{Wen1992, Wen1993,Eisenstein2004}. 
In the opposite limit of large interlayer separation, the two layers can be described by a compressible state of decoupled composite Fermi liquids (CFLs)~\cite{Halperin1993}. 
The intermediate layer separation, where the possible transition between the two phases occurs is not entirely understood. 
There exist many different possibilities: a potential phase transition~\cite{Shibata2006,Zhu2017}, a composite boson (CB) exciton condensate~\cite{Lian2018}, phase coexistence between CBs and CFs~\cite{Simon2003}, and several interlayer pairing instabilities~\cite{Morinari1999,Kim2001,Moller2008,Sodemann2017,Bonesteel1993,Bonesteel1996,Cipri2014,Isobe2017,Wagner2021}. 

More insight into the intermediate region is provided by two recent experiments~\cite{Eisenstein2019, Liu2022}. 
In an interlayer tunneling experiment, Eisenstein \textit{et. al.}~\cite{Eisenstein2019} report that the tunneling pseudogap in widely separated layers is suppressed at interlayer distance larger than what is expected for the transition to the XC phase. 
In a temperature dependent Coulomb drag and counter-flow experiment on QH bilayers of graphene, Liu \textit{et. al.}~\cite{Liu2022} report that a significant fraction of excitons are present at temperatures higher than the transition temperature $T_c$ associated with the formation of the XC phase. 
The main features of these experiments indicate a region of preformed pairs of electrons and holes which go over a smooth BCS-BEC-like crossover to the XC phase. 
The BCS-BEC crossover picture is also supported by a large overlap between numerical exact-diagonalization and a trial BCS wave function for interlayer CF electron-hole pairs~\cite{Wagner2021}.

\begin{figure}
\includegraphics[width=0.5\textwidth]{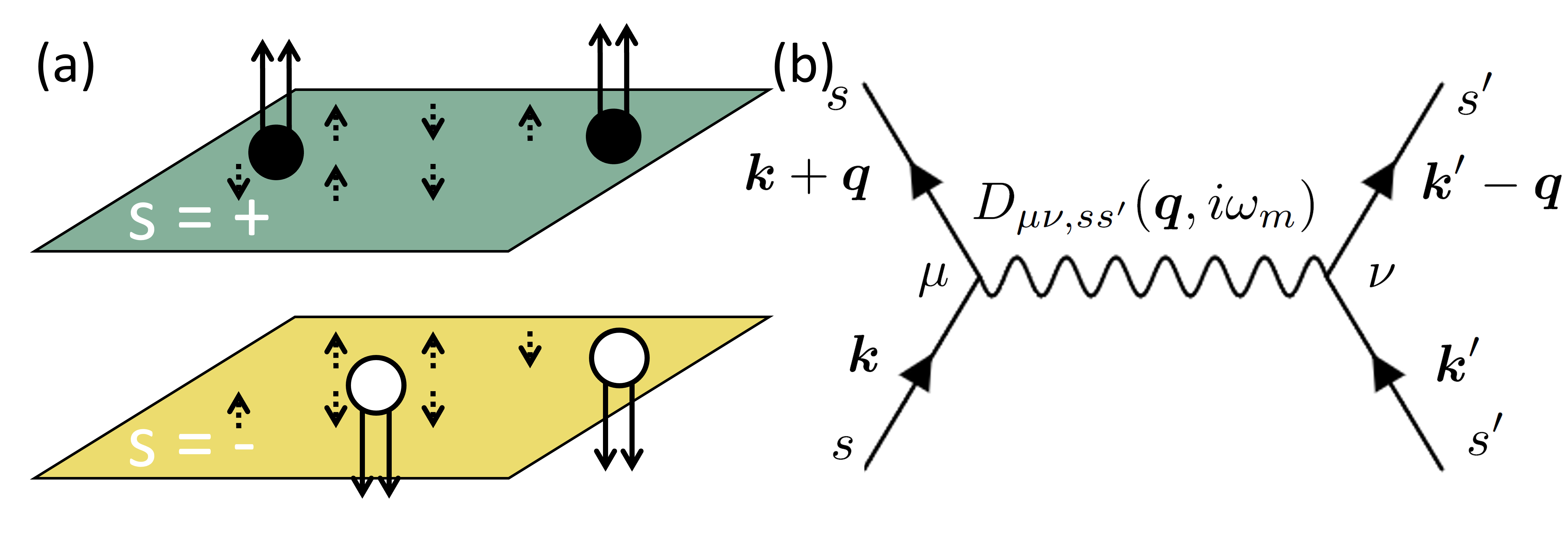}
\caption{(a) Schematic of the CEL-CHL bilayer system. The solid spheres with arrows are the CEs and the void spheres are CHs. The dashed arrows stand for the gauge fluctuations $\tilde a^s$. (b) Feynman diagram for the effective interaction between CEs and CHs mediated by the effective gauge propagator.}
\label{fig: schematic_main}
\end{figure}

In this Letter, we formulate a CS theory of a bilayer of CEL in one layer and CHL in the other as schematically shown in Fig.~\ref{fig: schematic_main} (a). 
Microscopically, the two layers are coupled by Coulomb interactions of bare particles. 
Starting from a large separation limit, within the random-phase-approximation (RPA)~\cite{Bohm1953,Fetter1991}, we arrive at an effective action for the fluctuations of the CS gauge fields around their mean-field value. 
The CS gauge fields mediate the effective interactions between the CFs in the two layers as shown in Fig.~\ref{fig: schematic_main} (b). 
This approach was previously used for the bilayers where both layers are a CEL~\cite{Bonesteel1993,Bonesteel1996,Cipri2014,Isobe2017}, and emergence of pairing instabilities between the two CELs was predicated. 
We show that in the CEL-CHL theory the effective interactions also lead to an interlayer pairing instability (\textit{albeit} in the CE-CH channel) and facilitate the formation of interlayer ``composite excitons (CX)". 
These CXs are thus expected to start forming at large layer separation and can be thought of as the particles that undergo the BCS-BEC crossover as the layers are brought closer. 
We also show that the dominant pairing occurs in the $s$-wave channel of the CXs, thus theoretically justifies the picture in Ref.~\cite{Wagner2021}. 
This is in contrast to the CEL-CEL theory where $p$-wave pairing instability is favoured~\cite{Moller2008,Moller2009,Isobe2017}. 
This difference is rooted in the density-current coupling, which in the CEL-CEL theory breaks time reversal symmetry (TRS) to favour a particular $p$-wave channel. 
In our CEL-CHL theory, the density-current coupling has opposite sign in the two layers and drops out in the Cooper channel, leading to an effective TRS.

Finally, we show that when the full frequency dependence of the effective interactions is taken into account, the $s$-wave CEL-CHL pair is more tightly bound than the $p$-wave CEL-CEL pair, thus favouring a BCS-BEC crossover picture in this theory.
Furthermore, when the two layers have slightly mismatched densities, \ie $\nu_T = (\frac{1}{2}-\delta) + (\frac{1}{2}+\delta)$, the two theories have different pair-breaking mechanisms, which also favours larger pair binding energies in the CEL-CHL theory. 
These can be possibly used in experiments to further confirm BCS-BEC crossover of CXs.

\sect{CS Theory} 
For the CEL-CHL CS theory, we first need to consider the theory of holes in the lowest Landau level (LLL). 
For our purpose, we consider the Lagrangian theory of holes in the LLL formulated in Ref.~\cite{Barkeshli2015} (after setting  $\hbar = c = e = 1$):
\begin{align}
\label{eq: hole lagrangian}
\begin{split}
    \mathcal{L}_{\text{hole}} = \, &\hd \left( i\partial_t - A_t + \mu_h \right) h - \frac{1}{2m_h} \hd \left( i \vec\nabla - \vec A \right)^2 h\\
    &- V(h) + \frac{1}{2\pi} A_t \vec\nabla \wedge \vec A,
\end{split}
\end{align}
with the Coulomb interaction potential
\begin{equation}
    V(h) = \frac{1}{2} \int d^2 \vec y \, \hd h \frac{1}{|\vec x - \vec y|} \hd h,
\end{equation}
where $h$ is the hole field, $e$ is the elementary charge, $A$ is the external vector potential, $\mu_h$ is the hole chemical potential, and $m_h$ is the hole effective mass. Importantly, the hole field has opposite charge in the external gauge field $A$ compared to the electron field. 
The CS term associated with the external field in the Lagrangian incorporates the effects of the filled LLL. 
Functional variation of the action with respect to the external field $A$ yields
\begin{align}
    n_e + n_h &= \frac{B}{2\pi},\\
    (\vec j_e)_i + (\vec j_h)_i &= \frac{\epsilon_{ij}}{2\pi} E_j,
\end{align}
where $\epsilon_{ij}$ is the Levi-Civita symbol. The first equation means that electrons plus holes fill the LLL. The second equation is the Hall conductivity. 
While, at the level of the Hamiltonian, the hole theory in the LLL can be obtained by a simple particle-hole transformation on the electron theory, in the Lagrangian incorporation of the extra CS term above puts important physical constraints. 
This extra CS term does not change the Hamiltonian~\cite{Lopez1991}.

We transform the holes $h$ into composite holes (CHs) $\chi$ by attaching two magnetic flux quanta to each [see Fig.~\ref{fig: schematic_main} (a)]. In the Lagrangian, this is achieved by including the dynamical gauge field $a$~\cite{Barkeshli2015}:
\begin{align}
\label{eq: CH lagrangian}
\begin{split}
    \mathcal{L}_{\text{CH}} = \, &\chid \left( i\partial_t + a_t - A_t + \mu_{\chi} \right) \chi - V(\chi)\\
    - \frac{1}{2m_{\chi}} &\chid \left( i \vec\nabla + \vec a - \vec A \right)^2 \chi + \frac{1}{2\pi} A_t \vec\nabla \wedge \vec A - \frac{1}{4\pi} a_t \vec\nabla \wedge \vec a,
\end{split}
\end{align}
where the interaction potential is the same as above since the density of holes and CHs is the same. The equation of motion of the dynamical gauge field stems from the second CS term and reads: $\vec\nabla \wedge \vec a = 4\pi \chid \chi$; which again has opposite sign compared to the case of the CF field $\psi$ and its associated dynamical gauge field: $\vec\nabla \wedge \vec a = -4\pi \psid \psi$. 
The Halperin-Lee-Read theory and hence the CF Lagrangian break the particle-hole symmetry~\cite{Halperin1993}. The CHL theory described by the above Lagrangian cannot be obtained by a simple particle-hole transformation of the CEL Lagrangian.  
Instead, as argued in Ref.~\cite{Barkeshli2015}, this CHL at $\nu =1/2$ is a topologically distinct state to the CEL at $\nu=1/2$. 
Pairing between these two topologically distinct CFLs is our main motivation. 

For our bilayer system, we take CFs as the degrees of freedom in one layer and CHs in the other; and accordingly denote $\psi_{+/-} = \psi / \chi$. The total Euclidean Lagrangian is~\cite{supplementary}
\begin{align}
\label{eq: euclidean lagrangian}
\begin{split}
    \mathcal{L} = \, & \sum_{s=+/-} \bigg[ \psid_s \left( \partial_{\tau} + ia^s_{\tau} + s iA_{\tau} - \mu_s \right) \psi_s\\
    &+ \frac{1}{2m_s} \psid_s \left( i \vec\nabla + \vec a^s + s \vec A \right)^2 \psi_s + \frac{si}{4\pi}  a^s_{\tau} \vec\nabla \wedge \vec a^s \\
    & + V(a^s) \bigg] + \frac{i}{2\pi} A_{\tau} \vec\nabla \wedge \vec A + V_{+-}(a^+, a^-),
\end{split}
\end{align}
with the interactions
\begin{align}
    V(a^s) &= \, \frac{1}{2} \int \frac{d^2 \vec y}{(4\pi)^2} \, \vec\nabla \wedge \vec a^s \frac{1}{|\vec x - \vec y|} \vec\nabla \wedge \vec a^s,\\
    V_{+-}(a^+, a^-) &= \int \frac{d^2 \vec y}{(4\pi)^2} \, \vec\nabla \wedge \vec a^+ \frac{1}{\sqrt{|\vec x - \vec y|^2 + d^2}} \vec\nabla \wedge \vec a^-.
\end{align}
In the expressions for the interactions above we already enforce the equations of motion for the dynamical gauge fields.

\sect{Effective Interaction}
The Lagrangian in Eq.~(\ref{eq: euclidean lagrangian}) possesses the mean-field solution $a^s = -s A$. We denote fluctuations in the dynamical gauge fields around this saddle-point as $\tilde a^s$. Furthermore, in the Coulomb gauge, the spatial part can be written as $\vec{\tilde a} (\vec q, i\omega_m) = \tilde a_1 (\vec q, i\omega_m) \hat z \wedge \vec{\hat q}$, where $\omega_m = 2\pi m T$ is a bosonic Matsubara frequency. Within the RPA, the effective action for the CS gauge field fluctuations up to second order is~\cite{Bonesteel1993,Bonesteel1996,Cipri2014,Isobe2017,supplementary}
\begin{align}
\label{eq: effective action}
\begin{split}
    S_{\text{eff}} = \, &\frac{T}{2} \sum_{\omega_m} \int \frac{d^2 \vec q}{(2\pi)^2} \sum_{\mu \nu, ss'}\\
    &\tilde a_{\mu}^s (\vec q, i\omega_m) D_{\mu \nu, ss'}^{-1} (q, i\omega_m) \tilde a_{\nu}^{s'} (-\vec q, -i\omega_m),
\end{split}
\end{align}
where $D (q,i\omega_m)$ is the gauge propagator and $\mu, \nu $ are temporal and spatial coordinates denoted by $0$ and $1$ respectively.  
We are interested in the low-energy, long-wavelength modes, \textit{i.e.} the regime $\frac{\omega_m}{\epsilon_F} \ll \frac{q^2}{k_F^2} \ll 1$ and $qd \ll 1$. The most singular terms in the gauge propagator are then~\cite{supplementary}
\begin{align}
\begin{split}
    &D_{11, ss}(q, i\omega_m)
    \approx -\frac{1}{\chi q^3} \left( \frac{m_+ m_-}{2\pi (4\pi)^2} q + \frac{m_+ m_- k_F}{(2\pi)^3} \frac{|\omega_m|}{q} \right),\\
    &D_{11, s(-s)}(q, i\omega_m) 
    \approx \frac{1}{\chi q^3} \frac{m_+ m_-}{2\pi (4\pi)^2} q,
\end{split}
\end{align}
where $\chi = \frac{m_+ m_- k_F}{2 (2\pi)^4} \frac{|\omega_m|}{q^3} + \left(\frac{m_+ + m_-}{192 \pi^4} + \frac{2d m_+m_-}{(4\pi)^4} \right)$. 
These are interpreted as current-current correlations. 
Importantly, the intra- and interlayer terms are equally singular.

After summing over the spatio-temporal indices, the CS gauge field  mediates an effective interaction within the CEL/CHL, and between the CEL and the CHL, given by their respective matrix elements ~\cite{Isobe2017,supplementary}
\begin{align}
\begin{split}
    V_{ss'}^{\text{eff}}(\vec k, \vec k', \vec q, i\omega_m) =
    -ss'\frac{(\vec{\hat q} \wedge \vec k) (\vec{\hat q} \wedge \vec k')}{2m_s m_{s'}}\\
    \times \left[ q^2 \left(\frac{m_+ + m_-}{6\pi m_+m_-} + \frac{d}{4\pi}\right) + \frac{|\omega_m|}{q} \frac{k_F}{\pi} \right]^{-1}.
\end{split}
\end{align}
The dominant contribution in the interlayer Cooper channel $V_c(\vec k, \vec q, i\omega_m) = V_{+-}^{\text{eff}}(\vec k, -\vec k - \vec q, \vec q, i\omega_m) + V_{-+}^{\text{eff}}(\vec k, -\vec k - \vec q, \vec q, i\omega_m)$ is attractive. 
Thus,to this point, the pairing instability in the CEL-CHL theory is similar to the one obtained for the CEL-CEL pairing theory~\cite{Bonesteel1993,Bonesteel1996,Cipri2014,Isobe2017}. 
Our starting theory in Eq.~(\ref{eq: euclidean lagrangian}) differs from the CEL-CEL theory, not only via the additional CS term of the external gauge field, but also regarding the charges of the CFs. 
However in the mean-field solution, the attached flux cancels the external field for the electrons and holes alike. 
The remaining fluctuations near the mean-field take both positive and negative values. 
Thus,the electrons and holes experience like charges in their respective CS gauge field fluctuations and resultant like charge currents. 
Since the dominant pairing contributions stem from the current-current coupling, to this level, there is yet no difference between CEL-CEL and CEL-CHL pairing. 
Further analysis shows that this attraction can be associated with the out-of-phase interlayer current fluctuations $ b_1^- = \frac{\tilde a^+_1 - \tilde a^-_1}{\sqrt{2}}$~\cite{supplementary}. 
As we will see below, the difference between the two theories appears in the density-current coupling of the CS gauge fields, which leads to different pairing symmetries.

\sect{Pairing Symmetry}
Within Eliashberg theory~\cite{Marsiglio2020}, the inverse Green's function in the Nambu space spanned by $(\psi_+(\vec k), \psi_-(\vec k), \psid_+(-\vec k), \psid_-(-\vec k))^\intercal$ is
\begin{align}
    G^{-1}(\vec k, i\epsilon_n) = 
    \begin{pmatrix}
        (i\epsilon_n Z_n^s - \xi_{\vec k}^s) \delta_{ss'} & \hat \phi_n(\vec k)\\
        \hat \phi_n(\vec k)^{\dagger} & (i\epsilon_n Z^s_n + \xi_{\vec k}^s) \delta_{ss'}
    \end{pmatrix},
\end{align}
with $Z_n^s$ the quasiparticle residue, $\hat \phi_n(\vec k)$ the anomalous self-energy, $\xi_{\vec k}^s = \frac{\vec k^2}{2m_s} - \mu_s$, and $\epsilon_n = 2\pi (n+1) T$ a fermionic Matsubara frequency. $Z_n^s$ and $\hat\phi_n(\vec k)$ get corrections from the exchange and Cooper channel interaction respectively. Going beyond the dominant current-current terms, these are written as~\cite{Isobe2017,supplementary}
\begin{align}
\begin{split}
    &V_{ex}(\vec k, \vec q, i\omega_m) = -\frac{1}{2} \sum_{\mu \nu, ss'}
    \begin{pmatrix}
    1 & -i \frac{\vec{\hat q} \wedge \vec k}{m_{s'}}\\
    i \frac{\vec{\hat q} \wedge \vec k}{m_s} & \frac{(\vec{\hat q} \wedge \vec k)^2}{m_s m_{s'}}
    \end{pmatrix}_{\mu \nu}\\
    &D_{\mu \nu, ss'}(q, i\omega_m) \left[ \delta_{s, +} \delta_{s', +} + \delta_{s, -} \delta_{s',- } \right],
\end{split}
\end{align}
and
\begin{align}
\begin{split}
    &V_c(\vec k, \vec q, i\omega_m) = \frac{1}{2} \sum_{\mu \nu, ss'}
    \begin{pmatrix}
    1 & i \frac{\vec{\hat q} \wedge \vec k}{m_{s'}}\\
    i \frac{\vec{\hat q} \wedge \vec k}{m_s} & -\frac{(\vec{\hat q} \wedge \vec k)^2}{m_s m_{s'}}
    \end{pmatrix}_{\mu \nu}\\
    &D_{\mu \nu, ss'}(q, i\omega_m) \left[ \delta_{s, +} \delta_{s', -} + \delta_{s, -} \delta_{s', +} \right].
\end{split}
\end{align}

The interlayer density-current interactions determine whether the $\pm l$ pair angular momentum states are degenerate as they are the only terms sensitive to the direction of the inserted flux quanta~\cite{Isobe2017,supplementary}. Their contribution to the Cooper channel
\begin{align}
\label{eq: off-diagonal terms Cooper channel}
\begin{split}
    \frac{i}{2} \vec{\hat q} \wedge \vec k \bigg[ \frac{D_{10,+-}(q, i\omega_m) + D_{01,-+}(q, i\omega_m)}{m_+}\\
    + \frac{D_{10,- +}(q, i\omega_m) + D_{01,+-}(q, i\omega_m)}{m_-} \bigg]
\end{split}
\end{align}
vanishes in the low-energy regime because the density-current couplings have opposite signs in the two layers~\cite{supplementary}. The density-current term originates from the CS term because it couples the spatial and the temporal components of the gauge field. 
Since the CEL and the CHL have the same charge with respect to their respective gauge fluctuations, they have the same charge current in response to the gauge fluctuations. 
However, since the CEL and the CHL have mean-field charge densities of opposite sign, the density-current correlators in the two layers have equal and opposite contributions which cancel to restore an effective TRS. 
This effective TRS ultimately leads to the degenerate $\pm l$ pairing channels.
This is in contrast to the CEL-CEL theory. 
In the CEL-CEL pairing scenario, since the charge-density as well as the charge currents in the two identical CFLs have the same sign, they add up to break the TRS~\cite{Isobe2017}.

Having shown that $\pm l$ channels are degenerate in the CEL-CHL pairing, we next analyse the relative strength of different $|l|$ channels. 
For this purpose, we make the effective mass approximation $m_+ = m_- = m^*$ which stands on good theoretical~\cite{Barkeshli2015} and experimental~\cite{Pan2020} footing. 
We expect our main results to be unchanged by small variations in the effective masses. 
Under this approximation, Eq.~(\ref{eq: off-diagonal terms Cooper channel}) vanishes for any momentum and frequency~\cite{supplementary}.
To assess the stability of the pairing we consider the effective coupling constants
\begin{align}
\label{eq: effective couplings}
\begin{split}
    \lambda_{Z,m} &= \int \frac{d^2 \vec q}{(2\pi)^2} \delta(\xi_{\vec k + \vec q}) V_{ex}(\vec k, \vec q, i\omega_m),\\
    \lambda_{\phi,m}^{(l)} &= \int \frac{d^2 \vec q}{(2\pi)^2} \delta(\xi_{\vec k + \vec q}) V_c(\vec k, \vec q, i\omega_m) \left( 1 + \frac{q}{k_F} e^{il(\theta_{\vec q} - \theta_{\vec k})} \right) ^l,
\end{split}
\end{align}
where $\theta_{\vec q}$ is the polar angle of $\vec q$, and we assumed that the pairing happens only at the Fermi surface, \ie $|\vec k| = k_F$. In the limit of zero frequency and long-wavelengths, the effective couplings are approximated by their singular terms~\cite{Isobe2017,supplementary}
\begin{align}
\begin{split}
    \lambda_{Z,0} &\approx \frac{m^*}{k_F} \int_0^{2k_F} \frac{dq}{(2\pi)^2} \left( \frac{k_F^2}{16 \pi^3 q^2 \chi_0} + \frac{\frac{k_F^2}{12 \pi^3 m^* \chi_0} - \frac{k_F^2 \chi'}{16 \pi ^3 \chi_0^2}}{q} \right),\\
    \lambda_{\phi,0}^{(l)} &\approx \frac{m^*}{k_F} \int_0^{2k_F} \frac{dq}{(2\pi)^2} \left( -\frac{k_F^2}{16 \pi ^3 q^2
   \chi_0} + \frac{\frac{d k_F^2}{16 \pi^3 \chi_0} + \frac{k_F^2 \chi'}{16 \pi^3 \chi_0^2}}{q} \right),
\end{split}
\end{align}
with $\chi_0 = \frac{m^*}{96 \pi^4} + \frac{2d (m^*)^2}{(4\pi)^4}$ and $\chi' = \frac{1}{144 \pi^4} - \frac{2(m^*)^2 d^2}{(4\pi)^4}$. Because the leading term in the anomalous self-energy is negative the CEL-CHL pairing is stable. This attractive interaction is mediated via the out-of-phase current fluctuations. The next order term corresponds to the repulsive in-phase current-current interactions~\cite{Isobe2017,supplementary}.

\begin{figure*}
\includegraphics[width=\textwidth]{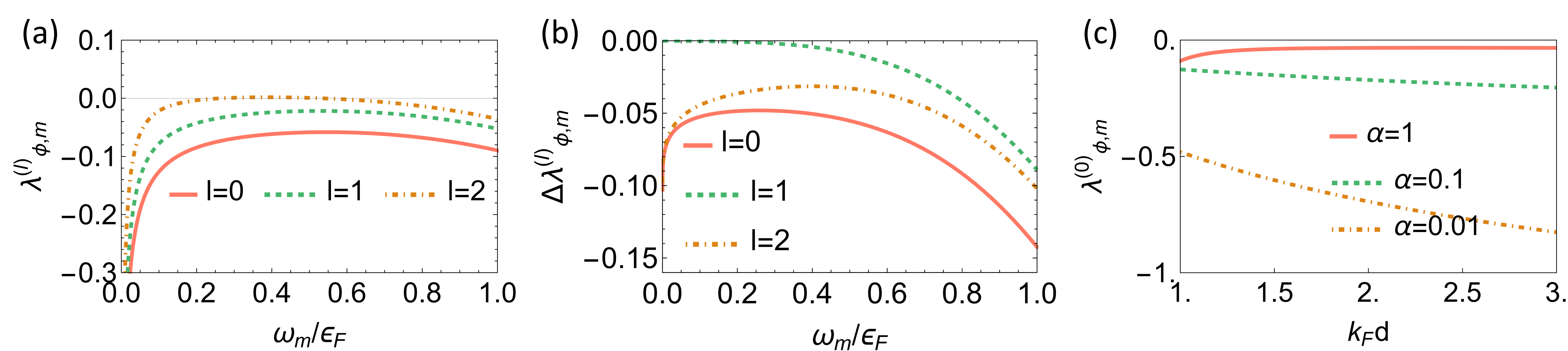}
\caption{(a) Effective couplings $\lambda_{\phi, m}^{(l)}$ as functions of the frequency. The integrals in Eq.~(\ref{eq: effective couplings}) diverge for small $q$ both for zero and finite frequencies. This is cured by introducing the cutoff $q_c = 10^{-5} k_F$. We set $k_F d = 1$. (b) Difference between the effective couplings for CEL-CEL and CEL-CHL pairing $\Delta \lambda_{\phi, m}^{(l)} = \lambda_{\phi, m}^{CEL-CHL (0)} - \lambda_{\phi, m}^{CEL-CEL (l)}$, with $k_F d = 1$. (c) Effective coupling $\lambda_{\phi, m}^{(0)}$ when varying the Fermi wavevector $k_F$ and keeping  distance $d$, effective mass $m^*$, and frequency $\omega_m$ constant parametrised by $\alpha = 2m^{\ast}\omega_md^2$.}
\label{fig: figure_main}
\end{figure*}

The terms distinguishing the different pairing channels enter only at $q^0$ order:
\begin{equation}
    \lambda_{\phi,0}^{(l)} - \lambda_{\phi,0}^{(0)} \approx \frac{m^*}{k_F} \int_0^{2k_F} \frac{dq}{(2\pi)^2} \, \frac{4l^2}{128 \pi^3 \chi_0}.
\end{equation}
It is immediately clear from this that the favoured pairing happens in the $s$-wave channel. This can be explained by the fact that the Cooper channel interactions are isotropic.
Using analytic expressions for the propagators~\cite{Cipri2014,Isobe2017} and numerically performing the integrals in Eq.~(\ref{eq: effective couplings})~\cite{supplementary}, we obtain the effective couplings [see Fig.~\ref{fig: figure_main} (a)]. It is obvious that $l = 0$ is the strongest pairing channel for all frequencies.

\sect{CEL-CHL vs. CEL-CEL Pairing}
Next, we compare the relative stability of the pairs in the two theories, \textit{i.e.} $s$-wave CEL-CHL pairs (CXs) to the CEL-CEL Cooper pairs. 
In Fig.~\ref{fig: figure_main} (b), we therefore consider the difference between the leading CEL-CHL coupling $\lambda_{\phi, m}^{CEL-CHL (0)}$ and the three strongest CEL-CEL couplings $\lambda_{\phi, m}^{CEL-CEL (l=0,1,2)}$. 
The CEL-CHL $s$-wave channel dominates at all frequencies. 
Thus,after performing the Matsubara frequency sum, the CEL-CHL pairing theory leads to CXs that are more tightly bound than the Cooper pairs in the CEL-CEL theory. 
This suggests that CXs are more likely to preform in a large region before the exciton condensation~\cite{Eisenstein2019,Liu2022}. 
Thus, this theory has stronger tendency to undergo BCS-BEC crossover than the CEL-CEL theory. 
The CEL-CEL theory possibly has stronger tendency towards a first order phase transition~\cite{Schliemann2001, Zou2010}. 
However, the contrary opinions also exist for the CEL-CEL theory~\cite{Sodemann2017}.
We mention in passing that at zero-frequency the coupling constants for the dominant $l=1$ CEL-CEL pairing channel are degenerate with the $l=0$ channel of CXs. 

We also compare the response of the two theories to a slight density mismatch in the two layers, while keeping the total filling factor fixed, \textit{i.e.} when the filling factors are $\nu_T = (1/2-\delta) + (1/2+\delta)$. 
Since each bare particle is attached to two fluxes, in each layer the external field is not exactly cancelled by the flux when it is moved away from half filling of the LLL. 
This remnant field is often interpreted to lead to a new set of Landau levels to explain the fractionalization (``$\Lambda$-levels" of the composite fermions)\cite{Jain1989}. 
Apart from the $\Lambda$-levels, the density imbalance also changes the Fermi wave vector of the CFL ($k_F = \sqrt{4\pi n}$)~\cite{Halperin1993}. 
In the following discussion, we assume that density imbalance is small enough that the individual layer is still away from a fractional state. 
Thus, we can ignore the effect of $\Lambda$-levels and in the isolated CFL, the change in the $k_F$ remains the only effect of density imbalance.
In the CEL-CEL theory, while one of the layers is depleted, the density in the other layer increases, as a result $k_F$ also decreases in one layer and increases in the other layer. 
This leads to pair-breaking phenomena similar to the Pauli-pair breaking in conventional superconductors, where the difference in the energy of the two opposite spin electrons that form the pairs leads to eventual suppression of the superconductivity when the magnetisation energy overcomes the condensation energy~\cite{Chandra1962,Clogston1962}.

In contrast, in the CEL-CHL theory, the layer with the increased electron density is described by the CHL~\cite{Barkeshli2015}. 
Thus, the CH and CE densities are equal~\cite{supplementary}.  
As a result, the density imbalance of bare particle leads to identical depletion of the two layers in the CF description. 
Thus, Pauli-like pair breaking phenomena are absent in this theory. 
However, since $k_F$ decreases with the density depletion, the effective coupling constants $\lambda_{\phi, m}^{(l)}$ usually decrease for small frequencies [see Fig.~\ref{fig: figure_main} (c)]. 
This can lead to a different mechanism for the suppression of CX formation with respect to the density imbalance.  
Notice in the CEL-CEL theory, this mechanism is also present along with the the Pauli pair breaking. 
Thus,the CXs are also more robust to small density imbalances compared to the CEL-CEL Cooper pairs. 
Thus, detailed experimental study with density-imbalanced states can be used to further verify CX formation.

\sect{Conclusions}
We have studied the pairing of CEL and CHL in the quantum Hall bilayer at $\nu_T = \frac{1}{2} + \frac{1}{2}$ by analysing fluctuations in the CS gauge field around the mean-field solution within the RPA. These fluctuations mediate interactions between the CEL and the CHL. In the interlayer Cooper channel, these interactions are attractive and lead to a stable CEL-CHL pairing with $l = 0$ symmetry due to an effective TRS. 
This yields a microscopic understanding for numerical results~\cite{Wagner2021}, where an $s$-wave CEL-CHL pairing trial wavefunction was found to have the best overlap with exact diagonalization results. 
The CXs in our theory are formed in the large layer separation limit, which appears to be in agreement with the experimental findings of preformed pairs before the transition to the XC~\cite{Eisenstein2019, Liu2022}. 
However, the pairs in our theory are pairs of CEs and CHs, which must be contrasted with the pairs of bare electrons and holes. 
It is the latter that condense to form an XC in closely separated bilayers. 
We leave this relation between the condensate of CXs and the bare  excitons, along with the theory of the full BCS-BEC crossover regime for future investigations.  

\begin{acknowledgements}
R.~J.~S and G.~C.~acknowledge funding from a New Investigator Award, EPSRC grant EP/W00187X/1. R.~J.~S also acknowledges funding from Trinity College, University of Cambridge.
L.~R. and R.~J.~S acknowledge funding provided by the Winton programme and the Schiff foundation.
\end{acknowledgements}

\bibliography{bibliography}

\clearpage
\onecolumngrid
\appendix

\renewcommand\thefigure{\thesection.\arabic{figure}}

\section{CEL-CHL CS Theory}
\setcounter{figure}{0}

We set $\hbar = c = e = 1$. The Schr\"odinger Lagrangian density for a spin-polarised electron field $c$ in a background gauge field $A$ is~\cite{Barkeshli2015}
\begin{align}
\begin{split}
    \mathcal{L}_{\text{el}} = \cd \left( i\partial_t + A_t + \mu_e \right) c - \frac{1}{2m_e} \cd \left( i \vec\nabla + \vec A \right)^2 c - V(c),
\end{split}
\end{align}
where the Coulomb interaction term is
\begin{equation}
    V(c) = \frac{1}{2} \int d^2 \vec y \, \cd c \frac{1}{|\vec x - \vec y|} \cd c.
\end{equation}
In the CEL picture, we take the CE field $\psi$ as the underlying degree of freedom. The transformed Lagrangian is~\cite{Zhang1989}
\begin{align}
\begin{split}
    \mathcal{L}_{\text{CEL}} = \psid \left( i\partial_t + a_t + A_t + \mu_{\psi} \right) \psi - \frac{1}{2m_{\psi}} \psid \left( i \vec\nabla + \vec a + \vec A \right)^2 \psi - V(\psi) + \frac{1}{4\pi} a_t \vec \nabla \wedge \vec a,
\end{split}
\end{align}
where $a$ is the dynamical gauge field. The CS term yields the equation of motion $\vec\nabla \wedge \vec a = - 4 \pi \psid \psi$, which is the attachment of the two flux quanta to the CEs.

The $\nu_T = 1$ charge-balanced bilayer CEL Lagrangian is
\begin{align}
\begin{split}
    \mathcal{L}_{\text{CEL-CEL}} = \sum_{s = + / -} \bigg[ \psid_s \left( i\partial_t + a_t^s + A_t + \mu_{\psi} \right) \psi_s - \frac{1}{2m_{\psi}} \psid_s \left( i \vec\nabla + \vec a^s + \vec A \right)^2 \psi_s - V(\psi_s) + \frac{1}{4\pi} a_t^s \vec \nabla \wedge \vec a^s \bigg] - V_{+-}(\psi_+, \psi_-).
\end{split}
\end{align}
The Coulomb interaction between layers separated by a distance $d$ is
\begin{equation}
    V_{+-}(\psi_+, \psi_-) = \int d^2 \vec y \, \psi^{\dagger}_+ \psi_+ \frac{1}{\sqrt{|\vec x - \vec y|^2 + d^2}} \psi^{\dagger}_{-} \psi_{-}.
\end{equation}
Both intra- and interlayer interactions can also be written in terms of the spatial parts of the dynamical gauge fields by enforcing their equations of motion
\begin{align}
    V(a^s) &= \, \frac{1}{2} \frac{1}{(4\pi)^2} \int d^2 \vec y \, \vec\nabla \wedge \vec a^s \frac{1}{|\vec x - \vec y|} \vec\nabla \wedge \vec a^s,\\
    V_{+ -}(a^+, a^-) &= \, \frac{1}{(4\pi)^2} \int d^2 \vec y \, \vec\nabla \wedge \vec a^+ \frac{1}{\sqrt{|\vec x - \vec y|^2 + d^2}} \vec\nabla \wedge \vec a^-.
\end{align}

We now want to describe the $-$ layer in terms of CHs. The non-interacting part is the same as in Eq.~(\ref{eq: CH lagrangian}) and was derived in Ref.~\cite{Barkeshli2015}. The intralayer interaction is the same as before. The interlayer interaction, however, obtains a minus sign when expressed in terms of the CE and CH fields, which cancels when expressing it in terms of the dynamical gauge fields:
\begin{align}
\begin{split}
    V_{+-}(\psi_+, \psi_-) &= \, \int d^2 \vec y \, \psid_+ \psi_+ \frac{-1}{\sqrt{|\vec x - \vec y|^2 + d^2}} \psid_- \psi_-,\\
    V_{+-}(a^+, a^-) &= \frac{1}{(4\pi)^2} \int d^2 \vec y \, \vec\nabla \wedge \vec a^+ \frac{1}{\sqrt{|\vec x - \vec y|^2 + d^2}} \vec\nabla \wedge \vec a^-.
\end{split}
\end{align}
Hence, the total CEL-CHL Lagrangian is
\begin{align}
\begin{split}
    \mathcal{L}_{\text{CEL-CHL}} = &\sum_{s=+/-} \bigg[ \psid_s \left( i\partial_t + a_t^s + s A_t + \mu_s \right) \psi_s - \frac{1}{2m_s} \psid_s \left( i \vec\nabla + \vec a^s + s \vec A \right)^2 \psi_s - V(a^s) + \frac{s}{4\pi} a_t^s \vec \nabla \wedge \vec a^s \bigg]\\
    & - V_{+-}(a^+, a^-) + \frac{1}{2\pi} A_t \vec\nabla \wedge \vec A.
\end{split}
\end{align}
In finite-temperature imaginary-time formalism, the above Lagrangian becomes ($\tau = it, \partial_t = i \partial_{\tau}, dt = -id\tau, A_t = -i A_{\tau}$)
\begin{align}
\begin{split}
    \mathcal{L}_{\text{CEL-CHL}} = &\sum_{s=+/-} \bigg[ \psid_s \left( \partial_{\tau} + ia_{\tau}^s + s iA_{\tau} - \mu_s \right) \psi_s + \frac{1}{2m_s} \psid_s \left( i \vec\nabla + \vec a^s + s \vec A \right)^2 \psi_s + V(a^s) + \frac{si}{4\pi} a_{\tau}^s \vec \nabla \wedge \vec a^s \bigg]\\
    & + V_{+-}(a^+, a^-) + \frac{i}{2\pi} A_{\tau} \vec\nabla \wedge \vec A.
\end{split}
\end{align}

\section{Saddle-Point Approximation and Effective Action}
\setcounter{figure}{0}

The CEL-CHL Lagrangian possesses a mean-field solution $a^s = -s A$ exactly analogous to the one for the monolayer CEL where the external flux is cancelled out. Going beyond this, we want to take fluctuations $\tilde a^s = a^s + s \times A$ around the saddle-point into account. The Lagrangian in terms of the fluctuations becomes
\begin{align}
\begin{split}
    \mathcal{L}_{\text{CEL-CHL}} = &\sum_{s=+/-} \bigg[ \psid_s \left( \partial_{\tau} + i\tilde a_{\tau}^s - \mu_s \right) \psi_s + \frac{1}{2m_s} \psid_s \left( i \vec\nabla + \vec{\tilde a^s} \right)^2 \psi_s + V(\tilde a^s) + \frac{si}{4\pi} \tilde a_{\tau}^s \vec \nabla \wedge \vec{\tilde a^s} \bigg] + V_{+-}(\tilde a^+, \tilde a^-),
\end{split}
\end{align}
where we could replace the dynamical gauge field in the interaction terms with its fluctuations because only terms quadratic in the fluctuations will contribute in the effective action (\cf Eq.~(\ref{eq: effective action})); similarly, only quadratic terms from the CS terms were kept.
We write our fields now as functions of momenta $\vec q$ and bosonic/fermionic Matsubara frequencies $\omega_m = 2m\pi T / \epsilon_n = 2(n+1)\pi T$. Assuming the Coulomb gauge $\vec\nabla \wedge \vec{\tilde a^s} = 0$, we can reduce the spatial components of the fluctuations of the dynamical gauge field to its transverse part $\tilde a_1$:
\begin{equation}
    \tilde a_1(\vec q, i\omega_m) = \vec{\hat q} \wedge \vec{\tilde a} (\vec q, i\omega_m), \quad \vec{\tilde a} (\vec q, i\omega_m) = \tilde a_1(\vec q, i\omega_m) \hat z \wedge \vec{\hat q}.
\end{equation}
We will denote the imaginary time component with the index $0$ from now on.

The Lagrangian in momentum-frequency space is
\begin{align}
\begin{split}
    \mathcal{L}_{\text{CEL-CHL}} = \, &\sum_{s=+/-} \bigg[ -\psid_s(\vec q, i\epsilon_n) i\epsilon \psi_s(\vec q, i\epsilon_n) + \psid_s(\vec k + \vec q, i\epsilon_n + i\omega_m) i\tilde a_0^s(\vec q, i\omega_m) \psi_s(\vec k, i\epsilon_n)\\
    &+ \frac{1}{2m_s} \psid_s(\vec q, i\epsilon_n) \vec q^2 \psi_s(\vec q, i\epsilon_n) - \frac{1}{m_s} \psid_s(\vec k + \vec q, i\epsilon_n + i\omega_m) \tilde a_1^s(\vec q, i\omega_m) \vec{\hat q} \wedge \vec k \psi_s(\vec k, i\epsilon_n)\\
    &-\frac{1}{2m_s} \psid_s(\vec k + \vec q - \vec{q'}, i\epsilon_n + i\omega_m - i\omega_m') \tilde a_1^s(\vec q, i\omega_m) \tilde a_1^s(-\vec{q'}, -i\omega_m') \vec{\hat q} \cdot \vec{\hat q'} \psi_s(\vec k, i\epsilon_n)\\
    &-\frac{1}{2} \frac{1}{(4\pi)^2} q^2 \tilde a_1^s(\vec q, i\omega_m) V_{ss}(q) \tilde a_1^s(-\vec q, -i\omega_m) - \frac{s}{4\pi} q \tilde a_0^s(\vec q, i\omega_m) \tilde a_1^s(-\vec q, -i\omega_m) \bigg]\\
    &- \frac{1}{(4\pi)^2} q^2 \tilde a_1^+(\vec q, i\omega_m) V_{+-}(q) \tilde a_1^-(-\vec q, -i\omega_m),
\end{split}
\end{align}
with the intralayer interaction $V_{++}(q) = V_{--}(q) = \frac{2\pi}{q}$, and the interlayer interaction $V_{+-}(q) = V_{-+}(q) = \frac{2\pi e^{-qd}}{q}$.

The Green's functions for the fermionic fields are
\begin{equation}
    G^s(q, i\epsilon_n) = \frac{1}{i\epsilon_n - \epsilon^s(q)},
\end{equation}
with $\epsilon^s(q) = \frac{q^2}{2m_s}$. The bare inverse gauge field propagator is
\begin{equation}
    D_{\mu \nu, ss'}^{(0)} (q, i\omega_m)^{-1} = 
    \begin{pmatrix}
    0 & -\frac{sq}{4\pi} \delta_{ss'}\\
    -\frac{sq}{4\pi} \delta_{ss'} & -\frac{q^2 V_{ss'}(q)}{(4\pi)^2}
    \end{pmatrix}_{\mu \nu}.
\end{equation}
The remaining terms in the Lagrangian generate vertices between the fermionic fields and the dynamical gauge field. They yield the the one-loop corrections
\begin{align}
    \Pi_{00, ss'} (q, i\omega_m) &= \braket{ a_0^s(\vec q, i\omega_m) a_0^{s'}(-\vec q, -i\omega_m)} = \Pi_{00}^s (q, i\omega_m) \delta_{ss'},\\
    \Pi_{11, ss'} (q, i\omega_m) &= \braket{ a_1^s(\vec q, i\omega_m) a_1^{s'}(-\vec q, -i\omega_m)} = \Pi_{11}^s (q, i\omega_m) \delta_{ss'}.
\end{align}
Since the vertices are exactly the same as in the CEL-CEL case, the polarisation diagrams are the same as in Refs.~\cite{Cipri2014,Isobe2017}. We calculate the full inverse gauge propagator within the RPA~\cite{Bohm1953,Fetter1991}:
\begin{align}
    D_{\mu \nu, ss'} (q, i\omega_m)^{-1} &= D_{\mu \nu, ss'}^{(0)} (q, i\omega_m)^{-1} - \Pi_{\mu \nu, ss'} (q, i\omega_m)\\
    &= 
    \begin{pmatrix}
    -\Pi_{00}^+(q, i\omega_m) & -\frac{q}{4\pi} & 0 & 0 \\
    -\frac{q}{4\pi} & -\Pi_{11}^+(q, i\omega_m) - \frac{q^2}{(4\pi)^2} V_{++}(q) & 0 & - \frac{q^2}{(4\pi)^2} V_{+-}(q)\\
     0 & 0 & -\Pi_{00}^-(q, i\omega_m) & \frac{q}{4\pi} \\
     0 &- \frac{q^2}{(4\pi)^2} V_{-+}(q) & \frac{q}{4\pi} & -\Pi_{11}^-(q, i\omega_m) - \frac{q^2}{(4\pi)^2} V_{--}(q) \\
    \end{pmatrix}_{\mu \nu, ss'}.
\end{align}

\section{Low-Energy Long-Wavelength Approximation Propagator}
\setcounter{figure}{0}
\label{app: low-energy long-wavelength approximation propagator}

Ultimately, we will be interested in the long-wavelength and low-energy modes of the system. These will be in the regime $\frac{\omega_m}{\epsilon_F} \ll \frac{q^2}{k_F^2} \ll 1$ and $qd \ll 1$. We can then approximate the polarisation diagrams as~\cite{Cipri2014,Isobe2017}
\begin{equation}
    \Pi_{00}^s (q, i\omega_m) \approx -\frac{m_s}{2\pi}, \quad \Pi_{11}^s (q, i\omega_m) \approx \chi_d^s q^2 - \frac{m_s}{2\pi} \frac{\omega_m^2}{q^2} + \frac{k_F}{2\pi} \frac{|\omega_m|}{q},
\end{equation}
where $\chi_d^s = \frac{1}{24 \pi m_s}$ is the diamagnetic susceptibility.

We calculate the propagator matrix as $D(q, i\omega_m) = \frac{1}{\text{det} \left[D(q, i\omega_m)^{-1} \right]} \text{adj} \left[D(q, i\omega_m)^{-1} \right]$. The determinant term is
\begin{align}
\begin{split}
    \text{det} \left[D(q, i\omega_m)^{-1} \right] &= \left(\frac{q^2}{(4\pi)^2} - \Pi_{00}^+(q, i\omega_m) \left[ \Pi_{11}^+(q, i\omega_m) + \frac{q^2}{(4\pi)^2} V_{++}(q) \right] \right)\\
    &\times \left(\frac{q^2}{(4\pi)^2} - \Pi_{00}^-(q, i\omega_m) \left[ \Pi_{11}^-(q, i\omega_m) + \frac{q^2}{(4\pi)^2} V_{++}(q) \right] \right)\\
    &- \Pi_{00}^+(q, i\omega_m) \Pi_{00}^-(q, i\omega_m) \frac{q^4}{(4\pi)^4} V_{+-}(q)^2\\
    &\approx \frac{m_+ m_- k_F}{2 (2\pi)^4} |\omega_m| + q^3 \left(\frac{m_+ + m_-}{192 \pi^4} + \frac{2d m_+ m_-}{(4\pi)^4} \right) + q^4 \left( \frac{1}{144 \pi^4} - \frac{2m_+ m_- d^2}{(4\pi)^4} \right)\\
    &= \chi q^3 + \chi' q^4,
\end{split}
\end{align}
where we defined
\begin{equation}
    \chi = \frac{m_+ m_- k_F}{2 (2\pi)^4} \frac{|\omega_m|}{q^3} + \left(\frac{m_+ + m_-}{192 \pi^4} + \frac{2d m_+m_-}{(4\pi)^4} \right), \quad \chi' = \frac{1}{144 \pi^4} - \frac{2m_+ m_- d^2}{(4\pi)^4}.
\end{equation}
The propagators in the different layer sectors are then
\begin{equation}
    D_{++}(q, i\omega_m) \approx \frac{1}{\text{det} \left[D(q, i\omega_m)^{-1} \right]}
    \begin{pmatrix}
    q^3 \left( \frac{2\pi}{(4\pi)^4} + \frac{1 + \frac{m_-}{m_+}}{24\pi (4\pi)^2} + \frac{m_- d}{(4\pi)^3} \right) + |\omega_m| \frac{m_- k_F}{16 \pi^3} & -\left( q^2 \frac{m_-}{(4\pi)^3} + q^3 \frac{1}{48\pi^3} + |\omega_m| \frac{m_- k_F}{16 \pi^3} \right)\\
    -\left( q^2 \frac{m_-}{(4\pi)^3} + q^3 \frac{1}{48\pi^3} + |\omega_m| \frac{m_- k_F}{16 \pi^3} \right) & -\left( q \frac{m_+ m_-}{2\pi (4\pi)^2} + q^2 \frac{m_+}{24 \pi^3} + \frac{|\omega_m|}{q} \frac{m_+ m_- k_F}{(2\pi)^3} \right)
    \end{pmatrix},
\end{equation}
\begin{equation}
    D_{+-}(q, i\omega_m) = D_{-+}(q, i\omega_m)^\intercal \approx \frac{1}{\text{det} \left[D(q, i\omega_m)^{-1} \right]}
    \begin{pmatrix}
    -q^3 \frac{1}{2(4\pi)^3} & q^2 \frac{m_-}{(4\pi)^3} - q^3 \frac{m_- d}{(4\pi)^3}\\
    -q^2 \frac{m_+}{(4\pi)^3} + q^3 \frac{m_+ d}{(4\pi)^3} & \frac{m_+ m_-}{2\pi (4\pi)^2} \left(q - q^2 d + q^3\frac{d^2}{2}\right)
    \end{pmatrix},
\end{equation}
\begin{equation}
    D_{--}(q, i\omega_m) \approx \frac{1}{\text{det} \left[D(q, i\omega_m)^{-1} \right]}
    \begin{pmatrix}
    q^3 \left( \frac{2\pi}{(4\pi)^4} + \frac{1 + \frac{m_+}{m_-}}{24\pi (4\pi)^2} + \frac{m_+ d}{(4\pi)^3} \right) + |\omega_m| \frac{m_+ k_F}{16 \pi^3} & q^2 \frac{m_+}{(4\pi)^3} + q^3 \frac{1}{48\pi^3} + |\omega_m| \frac{m_+ k_F}{16 \pi^3}\\
    q^2 \frac{m_+}{(4\pi)^3} + q^3 \frac{1}{48\pi^3} + |\omega_m| \frac{m_+ k_F}{16 \pi^3} & -\left( q \frac{m_+ m_-}{2\pi (4\pi)^2} + q^2 \frac{m_-}{24 \pi^3} + \frac{|\omega_m|}{q} \frac{m_+ m_- k_F}{(2\pi)^3} \right)
    \end{pmatrix}.
\end{equation}

\section{Block-Diagonal Form of the Propagator}
\setcounter{figure}{0}

Upon making the effective mass approximation $m_+ = m_- = m^*$, which results in $\Pi_{\mu \nu}^+ = \Pi_{\mu \nu}^- = \Pi_{\mu \nu}$, there exists a different basis for the CS gauge fields in which the propagator becomes block-diagonal. We define
\begin{align*}
    b_0^+ &= \frac{1}{\sqrt{2}} (\tilde a^+_0 - \tilde a^-_0), \quad b_1^+ = \frac{1}{\sqrt{2}} (\tilde a^+_1 + \tilde a^-_1),\\
    b_0^- &= \frac{1}{\sqrt{2}} (\tilde a^+_0 + \tilde a^-_0), \quad b_1^- = \frac{1}{\sqrt{2}} (\tilde a^+_1 - \tilde a^-_1).
\end{align*}
The inverse propagator in this basis becomes
\begin{equation}
    D(q, i\omega_m)^{-1} = 
    \begin{pmatrix}
    D^+(q, i\omega_m)^{-1} & 0\\
    0 & D^-(q, i\omega_m)^{-1}
    \end{pmatrix},
\end{equation}
with
\begin{equation}
    D^+(q, i\omega_m)^{-1} = 
    \begin{pmatrix}
    -\Pi_{00}(q, i\omega_m) & -\frac{q}{4\pi}\\
    -\frac{q}{4\pi} & -\Pi_{11}(q, i\omega_m) - \frac{q^2}{(4\pi)^2} V_{++}(q) - \frac{q^2}{(4\pi)^2} V_{+-}(q)
    \end{pmatrix},
\end{equation}
\begin{equation}
    D^-(q, i\omega_m)^{-1} = 
    \begin{pmatrix}
    -\Pi_{00}(q, i\omega_m) & -\frac{q}{4\pi}\\
    -\frac{q}{4\pi} & -\Pi_{11}(q, i\omega_m) - \frac{q^2}{(4\pi)^2} V_{++}(q) + \frac{q^2}{(4\pi)^2} V_{+-}(q)
    \end{pmatrix}.
\end{equation}
We can invert these as
\begin{equation}
    D^{\pm}(q, i\omega_m) = -\frac{1}{\text{det}[D^{\pm}(q, i\omega_m)^{-1}]}
    \begin{pmatrix}
     \Pi_{11}(q, i\omega_m) + \frac{q^2}{(4\pi)^2} V_{++}(q) \pm \frac{q^2}{(4\pi)^2} V_{+-}(q) & -\frac{q}{4\pi}\\
    -\frac{q}{4\pi} & \Pi_{00}(q, i\omega_m)
    \end{pmatrix},
\end{equation}
with
\begin{equation}
    \text{det}[D^{\pm}(q, i\omega_m)^{-1}] = \Pi_{00}(q, i\omega_m) \left[\Pi_{11}(q, i\omega_m) + \frac{q^2}{(4\pi)^2} V_{++}(q) \pm \frac{q^2}{(4\pi)^2} V_{+-}(q)\right] - \frac{q^2}{(4\pi)^2}.
\end{equation}
Up to the $-$ sign in the off-diagonal terms, these are exactly the in-phase/out-of-phase propagators derived in Ref.~\cite{Isobe2017}. However, we have used a different basis and hence when transforming back into the CE-CH basis, the full propagator and the physical results will have crucial differences.

\section{Effective Interaction}
\setcounter{figure}{0}

The effective CS propagator mediates an interaction between the CEs and CHs. We define
\begin{equation}
    \mathcal{V} = \frac{1}{2} \sum_{ss'} \psid_s(\vec k + \vec q, i\epsilon_n + i\omega_m) \psid_{s'}(\vec k' - \vec q, i\epsilon_n' - i\omega_m) V_{ss'}^{\text{eff}}(\vec k, \vec k', \vec q, i\omega_m) \psi_{s'}(\vec k', i\epsilon_n') \psi_s(\vec k, i\epsilon_n),
\end{equation}
and hence
\begin{equation}
    V_{ss'}^{\text{eff}}(\vec k, \vec k', \vec q, i\omega_m) = \sum_{\mu \nu} M_{\mu \nu, ss'}(\vec k, \vec k', \vec q) D_{\mu \nu, ss'}(\vec q, i\omega_m),
\end{equation}
with
\begin{equation}
    M_{\mu \nu, ss'}(\vec k, \vec k', \vec q) = \frac{1}{2}
    \begin{pmatrix}
    1 & -i \frac{\vec{\hat q} \wedge \vec k'}{m_{s'}}\\
    i \frac{\vec{\hat q} \wedge \vec k}{m_s} & \frac{(\vec{\hat q} \wedge \vec k) (\vec{\hat q} \wedge \vec k')}{m_s m_{s'}}
    \end{pmatrix}.
\end{equation}

If we only consider the most singular terms in $q$, which are $D_{11, ss'}(q, i\omega_m)$, the effective interaction takes the form
\begin{equation}
    V_{ss'}^{\text{eff}}(\vec k, \vec k', \vec q, i\omega_m) = -ss'
    \frac{(\vec{\hat q} \wedge \vec k) (\vec{\hat q} \wedge \vec k')}{2m_s m_{s'}} \left[ q^2 \left(\frac{m_+ + m_-}{6\pi m_+ m_-} + \frac{d}{4\pi}\right) + \frac{|\omega_m|}{q} \frac{k_F}{\pi} \right]^{-1}.
\end{equation}

\section{Effective Couplings}
\setcounter{figure}{0}

Just as in the main text we will adopt the effective mass approximation from now on: $m_+ = m_- = m^*$.

Within Eliashberg theory~\cite{Isobe2017, Marsiglio2020}, the effective couplings for the quasiparticle residue $Z_n$ and the anomalous self-energy $\hat \phi_n(\vec k)$ are defined as
\begin{align}
\label{eqapp: effective couplings}
\begin{split}
    \lambda_{Z,m} &= \int \frac{d^2 \vec q}{(2\pi)^2} \delta(\xi_{\vec k + \vec q}) V_{ex}(\vec k, \vec q, i\omega_m),\\
    \lambda_{\phi,m}^{(l)} &= \int \frac{d^2 \vec q}{(2\pi)^2} \delta(\xi_{\vec k + \vec q}) V_c(\vec k, \vec q, i\omega_m) \left( 1 + \frac{q}{k_F} e^{il(\theta_{\vec q} - \theta_{\vec k})} \right) ^l,
\end{split}
\end{align}
with the exchange and Cooper channel interactions
\begin{align}
\begin{split}
    V_{ex}(\vec k, \vec q, i\omega_m) &= V_{++}^{\text{eff}}(\vec k, \vec k + \vec q, \vec q, i\omega_m) + V_{--}^{\text{eff}}(\vec k, \vec k + \vec q, \vec q, i\omega_m)\\
    &= -\frac{1}{2} \sum_{\mu \nu, ss'}
    \begin{pmatrix}
    1 & -i \frac{\vec{\hat q} \wedge \vec k}{m_{s'}}\\
    i \frac{\vec{\hat q} \wedge \vec k}{m_s} & \frac{(\vec{\hat q} \wedge \vec k)^2}{m_s m_{s'}}
    \end{pmatrix}_{\mu \nu} D_{\mu \nu, ss'}(q, i\omega_m) \left[ \delta_{s, +} \delta_{s', +} + \delta_{s, -} \delta_{s',- } \right],
\end{split}
\end{align}
\begin{align}
\begin{split}
    V_c(\vec k, \vec q, i\omega_m) &= V_{+-}^{\text{eff}}(\vec k, -\vec k - \vec q, \vec q, i\omega_m) + V_{-+}^{\text{eff}}(\vec k, -\vec k - \vec q, \vec q, i\omega_m)\\
    &= \frac{1}{2} \sum_{\mu \nu, ss'}
    \begin{pmatrix}
    1 & i \frac{\vec{\hat q} \wedge \vec k}{m_{s'}}\\
    i \frac{\vec{\hat q} \wedge \vec k}{m_s} & -\frac{(\vec{\hat q} \wedge \vec k)^2}{m_s m_{s'}}
    \end{pmatrix}_{\mu \nu} D_{\mu \nu, ss'}(q, i\omega_m) \left[ \delta_{s, +} \delta_{s', -} + \delta_{s, -} \delta_{s', +} \right].
\end{split}
\end{align}

Assuming that the gap is much larger than the Fermi energy, the pairing happens only on the Fermi surface. We can then perform the angular integration in Eq.~(\ref{eqapp: effective couplings}) and obtain~\cite{Isobe2017}
\begin{align}
\label{eqapp: lambda_z}
\begin{split}
    \lambda_{Z,m} = \frac{1}{(2\pi)^2} \frac{m^*}{k_F} \int_0^{k_F} dq \bigg[ &-\frac{1}{\sqrt{1 - \left(\frac{q}{k_F}\right)^2}} \Big( D_{00,++}(q, i\omega_m) + D_{00,--}(q, i\omega_m) \Big)\\
    &-\frac{k_F^2}{m^{*2}} \sqrt{1 - \left(\frac{q}{k_F}\right)^2} \Big( D_{11,++}(q, i\omega_m) + D_{11,--}(q, i\omega_m) \Big) \bigg],
\end{split}
\end{align}
and
\begin{align}
\label{eqapp: lambda_phi}
\begin{split}
    \lambda_{\phi,m}^{(l)} = \frac{1}{(2\pi)^2} \frac{m^*}{k_F} \int_0^{k_F} dq \bigg[ &\frac{1}{\sqrt{1 - \left(\frac{q}{k_F}\right)^2}} \cos\left(2l\arcsin \frac{q}{k_F}\right) \Big( D_{00,+-}(q, i\omega_m) + D_{00,-+}(q, i\omega_m) \Big)\\
    &-\frac{k_F^2}{m^{*2}} \sqrt{1 - \left(\frac{q}{k_F}\right)^2} \cos\left(2l\arcsin \frac{q}{k_F}\right) \Big( D_{11,+-}(q, i\omega_m) + D_{11,-+}(q, i\omega_m) \Big) \bigg].
\end{split}
\end{align}
Contrary to the CEL-CEL case in Ref.~\cite{Isobe2017}, for the CEL-CHL there are no odd-l terms in Eq.~(\ref{eqapp: lambda_phi}) because
\begin{equation}
    D_{10,+-}(q, i\omega_m) + D_{01,-+}(q, i\omega_m) + D_{10,- +}(q, i\omega_m) + D_{01,+-}(q, i\omega_m) = 0.
\end{equation}
This immediately tells us that the pairing channels $\pm l$ are degenerate.

\section{Low-Energy Long-Wavelength Approximation Effective Couplings}
\setcounter{figure}{0}

The pairing stability can be assessed by looking at the effective couplings at $\omega_m = 0$, where the low-energy long-wavelength approximation $\frac{\omega_m}{\epsilon_F} \ll \frac{q^2}{k_F^2} \ll 1$ and $qd \ll 1$ is valid. We use the approximate forms of the propagator entries from App.~\ref{app: low-energy long-wavelength approximation propagator} to write the effective couplings as
\begin{align}
\label{eqapp: lambda_z integral}
\begin{split}
    \lambda_{Z,0} = \frac{1}{(2\pi)^2} \frac{m^*}{k_F} \int_0^{2k_F} dq \Bigg[ &\frac{k_F^2}{16 \pi^3 q^2 \chi} + \frac{\frac{k_F^2}{12 \pi^3 m^* \chi} - \frac{k_F^2 \chi'}{16 \pi^3 \chi^2}}{q}\\
    &-\frac{2 \chi''}{\chi} - \frac{k_F^2 \chi'}{12 \pi^3
   m^* \chi^2} + \frac{k_F^2 \chi'^2}{16 \pi^3 \chi^3} - \frac{1}{128 \pi^3 \chi} + O(q) \Bigg],
\end{split}
\end{align}
and
\begin{align}
\label{eqapp: lambda_phi integral}
\begin{split}
    \lambda_{\phi,0}^{(l)} = \frac{1}{(2\pi)^2} \frac{m^*}{k_F} \int_0^{2k_F} dq \Bigg[ &-\frac{k_F^2}{16 \pi^3 q^2
   \chi} + \frac{\frac{d k_F^2}{16 \pi^3 \chi} + \frac{k_F^2 \chi'}{16 \pi^3 \chi^2}}{q}\\
   &- \frac{\frac{d^2 k_F^2}{32 \pi^3} + \frac{-4 l^2-1}{128 \pi^3}}{\chi} - \frac{d k_F^2 \chi'}{16 \pi^3 \chi^2} - \frac{k_F^2 \chi'^2}{16 \pi^3 \chi^3} - \frac{1}{64 \pi^3 \chi}  + O(q) \Bigg],
\end{split}
\end{align}
with
\begin{equation}
    \chi'' = \frac{5}{6 (4\pi)^3} + \frac{m^* d}{(4\pi)^3}.
\end{equation}
For $\omega_m = 0$, $\chi$ goes to $\chi_0 = \frac{m^*}{96 \pi^4} + \frac{2d (m^*)^2}{(4\pi)^4}$.

\section{Numerical Integration Effective Couplings}
\setcounter{figure}{0}

The integrals in Eqs. (\ref{eqapp: lambda_z integral},~\ref{eqapp: lambda_phi integral}) suffer from divergences as $q \rightarrow 0$. In Ref.~\cite{Isobe2017} it was shown that it is reasonable to introduce a cutoff $q_c = 10^{-5} k_F$. To perform the integrals numerically, we use the exact expressions for the polarisation diagrams $\Pi_{00}(q, i\omega_m), \Pi_{11}(q, i\omega_m)$ calculated in Ref.~\cite{Cipri2014}:
\begin{align}
\begin{split}
    \Pi_{00}(q, i\omega_m) &= F_1(q, i\omega_m) + F_1(q, -i\omega_m),\\
    F_1(q, i\omega_m) &= 2m^* f_1 \left(\frac{2q}{k_F}, \frac{i\omega_m}{\epsilon_F} - \frac{q^2}{k_F^2} \right),\\
    f_1(y,z) &= \frac{z}{2\pi y^2} \left( 1 - \left(1 - \frac{y^2}{z^2}\right)^{\frac{1}{2}} \right),
\end{split}
\end{align}
and
\begin{align}
\begin{split}
    \Pi_{11}(q, i\omega_m) &= F_2(q, i\omega_m) + F_2(q, -i\omega_m) + \frac{\epsilon_F}{2\pi},\\
    F_2(q, i\omega_m) &= 4\epsilon_F f_2 \left(\frac{2q}{k_F}, \frac{i\omega_m}{\epsilon_F} - \frac{q^2}{k_F^2} \right),\\
    f_2(y,z) &= \frac{z}{4\pi y^2} \left( 1 - \frac{2 z^2}{3 y^2} \left(1 - \left( 1 - \frac{y^2}{z^2}\right)^{\frac{3}{2}} \right) \right).
\end{split}
\end{align}
The two polarisation functions are real, unless analytically continued.
\begin{figure}[h]
    \centering
    \includegraphics[width=0.45\textwidth]{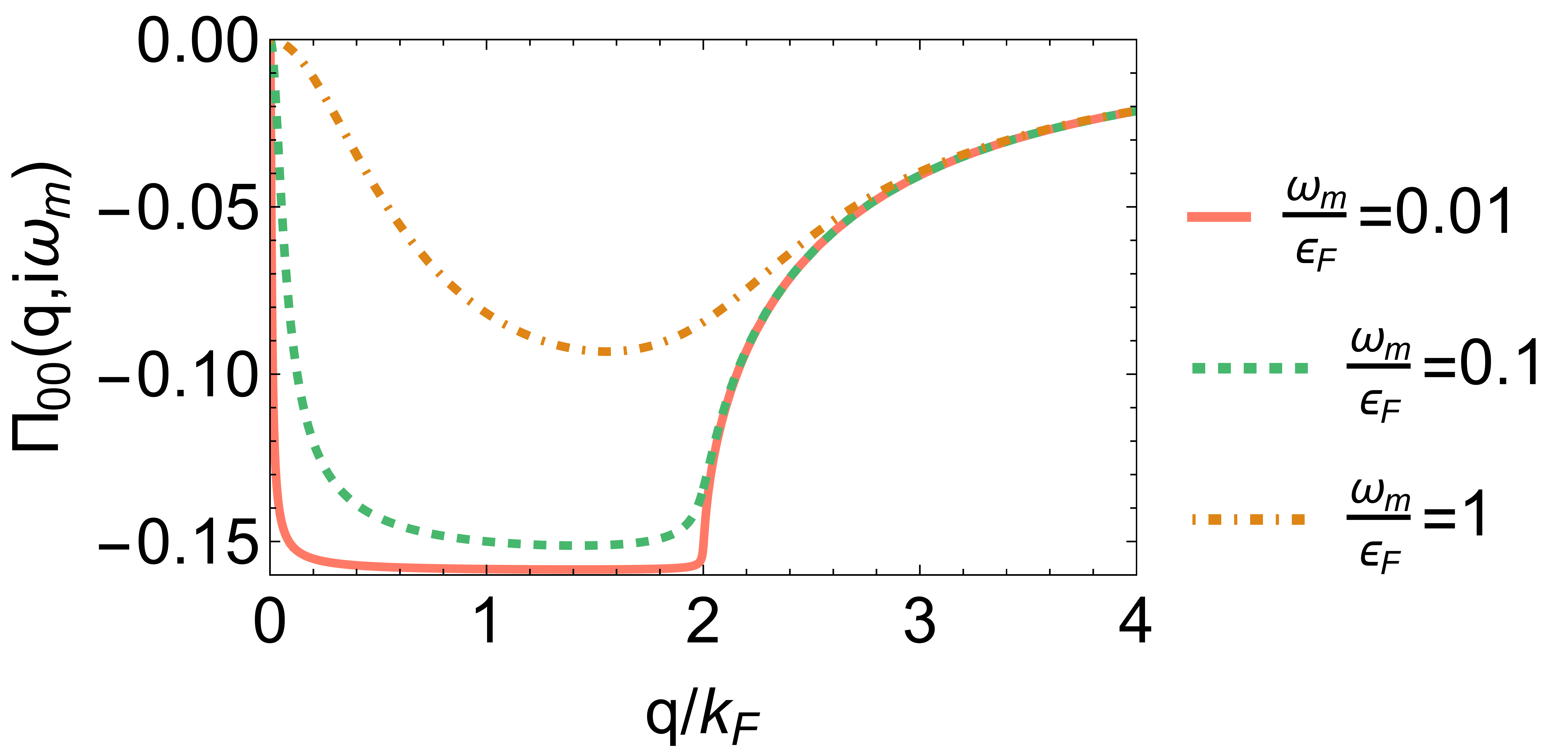}
    \quad
    \includegraphics[width=0.45\textwidth]{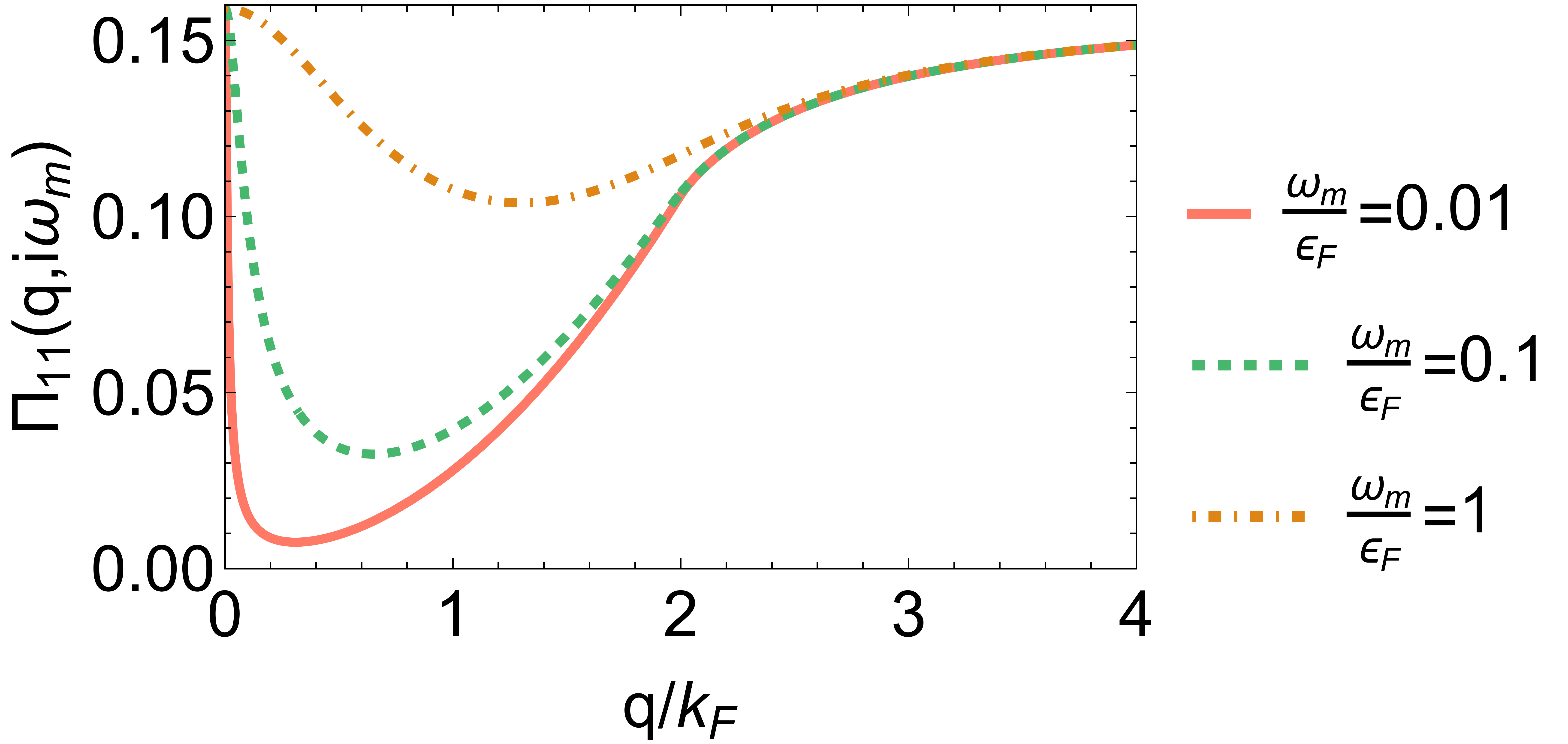}
    \caption{The two polarisation diagrams $\Pi_{00}(q, i\omega_m)$ and $\Pi_{11}(q, i\omega_m)$.}
    \label{figapp: polarization functions}
\end{figure}

\begin{figure}[h]
    \centering
    \includegraphics[width=0.4\textwidth]{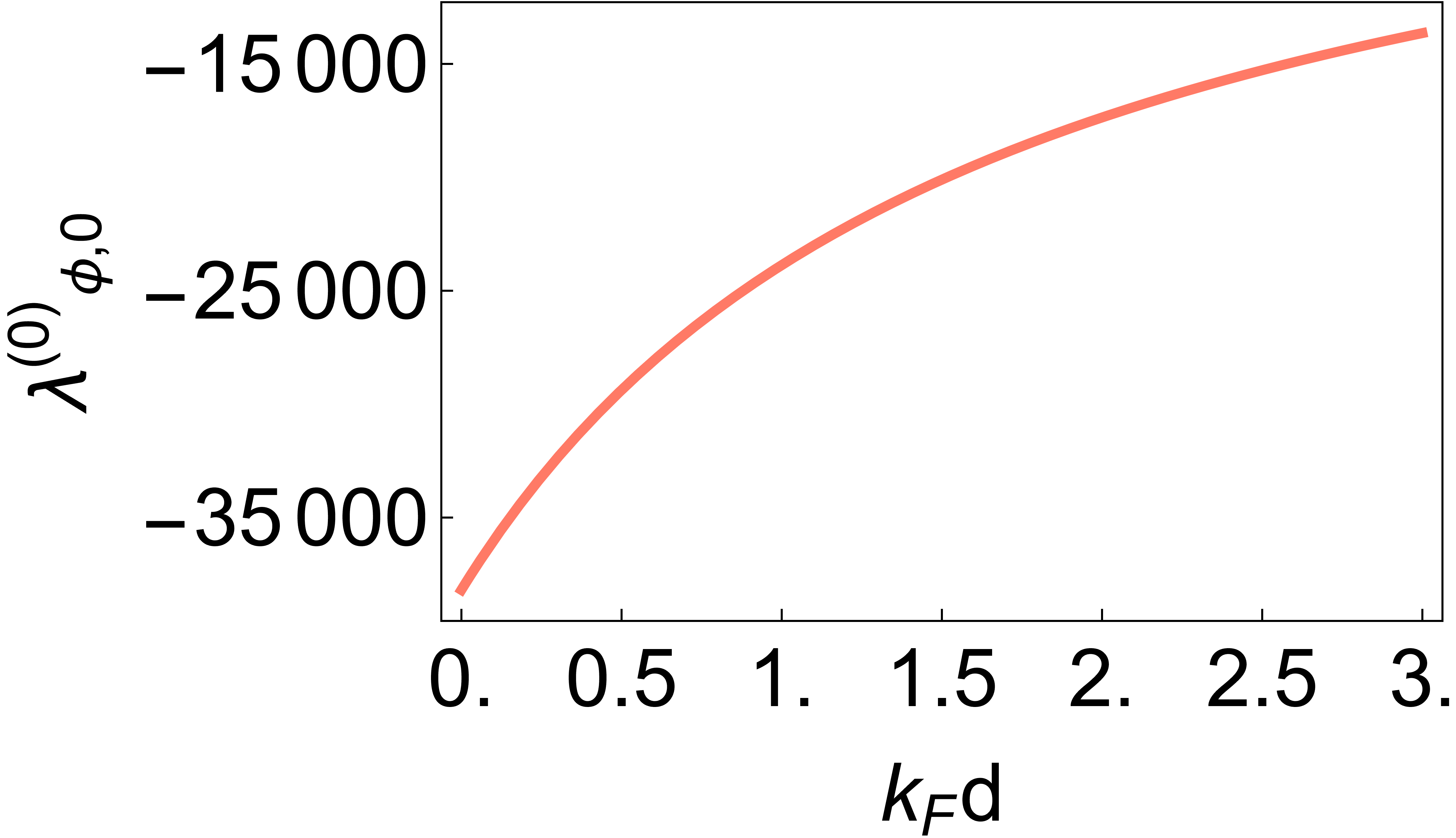}
    \quad
    \includegraphics[width=0.4\textwidth]{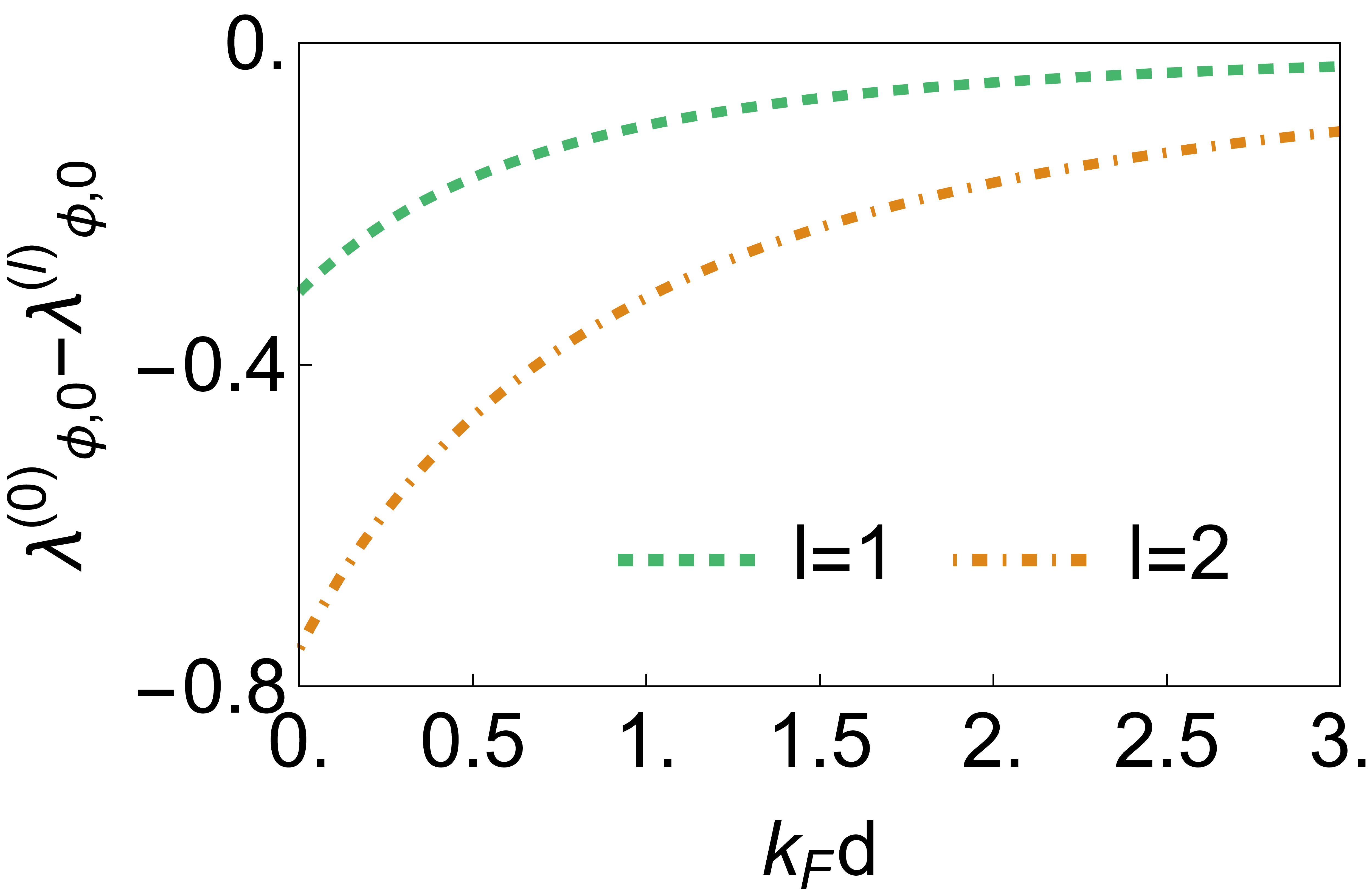}
    \caption{The anomalous self-energy $\lambda_{\phi,0}^{(0)}$ and the relative anomalous self-energy $\lambda_{\phi,0}^{(0)} - \lambda_{\phi,0}^{(l)}$ with cutoff $q_c = 10^{-5}$.}
    \label{figapp: lambda_phi_d}
\end{figure}

Now we are able to perform the integrals in Eqs. (\ref{eqapp: lambda_z},~\ref{eqapp: lambda_phi}). The frequency spectra are in Fig.~\ref{figapp: lambda_phi lambda_z} below.
\begin{figure}[h]
    \centering
    \includegraphics[width=0.4\textwidth]{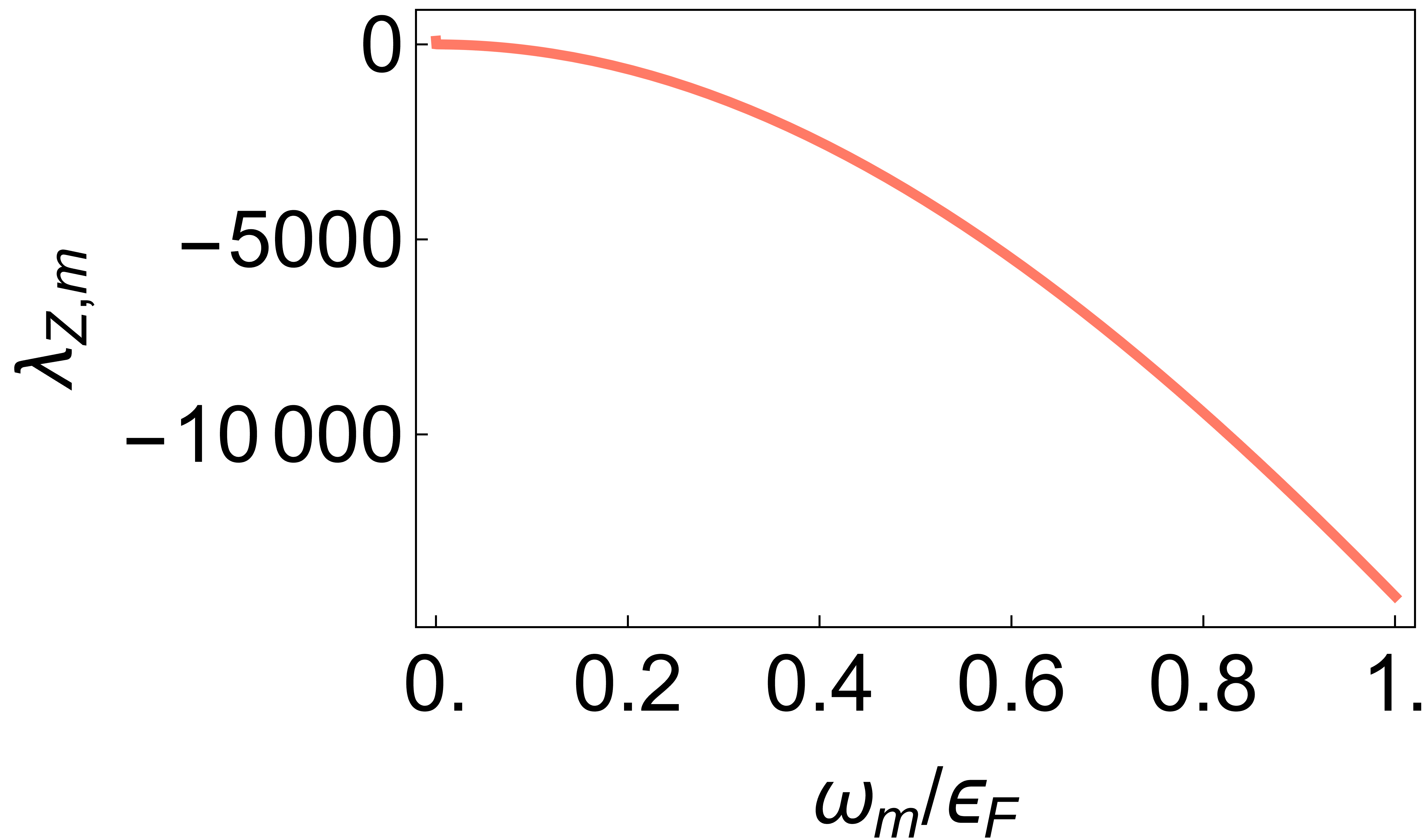}
    \quad
    \includegraphics[width=0.4\textwidth]{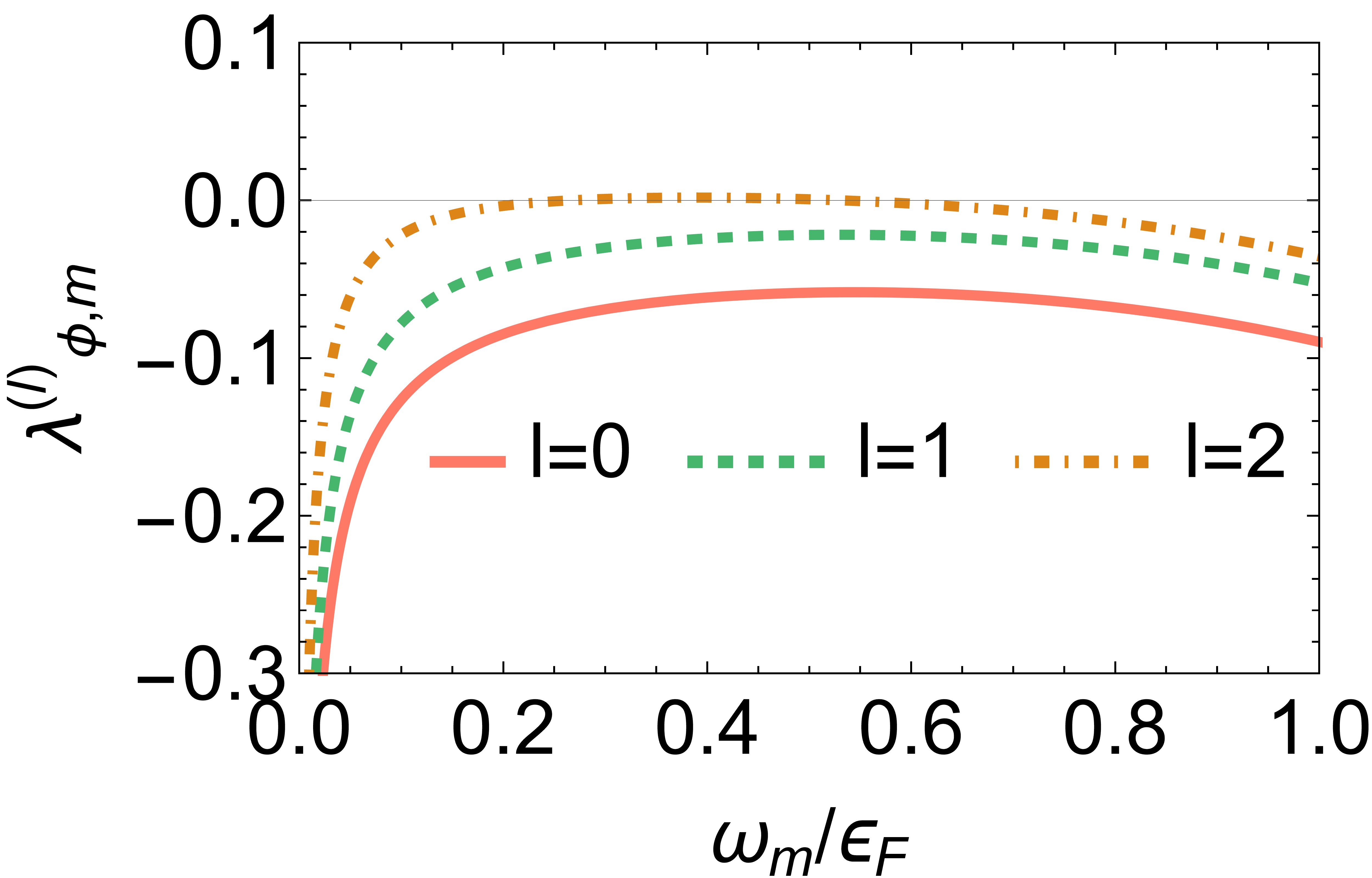}
    \caption{The quasiparticle residue $\lambda_{Z,m}$ and the anomalous self-energy $\lambda_{\phi,m}^{(l)}$ for $k_F d = 1$ and cutoff $q_c = 10^{-5}$.}
    \label{figapp: lambda_phi lambda_z}
\end{figure}
The behaviour of the anomalous self-energy $\lambda_{\phi,0}^{(l)}$ as a function of the layer separation is in absolute values for the $l = 0$ channel and in relative values for $l = 1,2$ in Fig.~\ref{figapp: lambda_phi_d}. We see that the pairing strength decreases with increasing layer separation and that the ordering of the different channels remains unchanged.

\section{Density Imbalance}
\setcounter{figure}{0}

As can be seen from the Lagrangian in Eq.~(\ref{eq: euclidean lagrangian}), the effective magnetic fields in the two layers are (taking the charge into account)
\begin{align}
    B_{\text{eff}}^+ &= B - 4 \pi n_+,\\
    B_{\text{eff}}^- &= -B + 4 \pi n_-.
\end{align}
Using the relation $n_+ + n_- = \frac{B}{2\pi}$, we notice that the CEs and CHs experience exactly the same effective magnetic field~\cite{Barkeshli2015}.

Introducing a charge imbalance $\nu_T = (1/2-\delta) + (1/2+\delta)$ between the two layers, we obtain the coupled Lagrangian near the mean-field solution
\begin{align}
\begin{split}
    \mathcal{L}_{\text{CEL-CHL}} = &\sum_{s=+/-} \bigg[ \psid_s \left( \partial_{\tau} + i\tilde a_{\tau}^s + is d A_{\tau} - \mu_s \right) \psi_s + \frac{1}{2m_s} \psid_s \left( i \vec\nabla + \vec{\tilde a}^s + s d\vec A \right)^2 \psi_s + V(\tilde a^s) + \frac{si}{4\pi} \tilde a_{\tau}^s \vec \nabla \wedge \vec{\tilde a}^s \bigg]\\
    & + V_{+-}(\tilde a^+, \tilde a^-),
\end{split}
\end{align}
where $d A_{\tau}$ denotes the remnant external field that is not cancelled by the flux attachment. 
This remnant field makes $B_{\text{eff}}$ in the two layers non-zero and pointing in opposite directions. This is true regardless of whether we describe the layers in terms of CEs or CHs. 
The additional effect from the density imbalance in the CEL-CHL Lagrangian appears from the last term above. 
However, since it is linear in the gauge fluctuations, it does not effect the theory and we can omit it from the Lagrangian. 
Thus, so far the charge imbalance has the same effect in the CEL-CEL as in the CEL-CHL. However, in the CEL-CHL theory, we should use CEs for the $\nu = 1/2 - \delta$ layer and CHs for the $\nu = 1/2 + \delta$ layer~\cite{Barkeshli2015}. This means that the Fermi surface shrinks in each of the two layers by the same amount. This is not the case in the CEL-CEL where the Fermi surface shrinks in one layer and expands in the other.

\end{document}